\documentstyle[aps,12pt]{revtex}

\begin{document}

\begin{titlepage}
\begin{flushright}
PUPT-1812\\

hep-th/9809057
\end{flushright}

\vspace{7 mm}

\begin{center}
{\huge The wall of the cave}
\end{center}
\vspace{10 mm}
\begin{center}
{\large  

A.M.~Polyakov\\
}
\vspace{3mm}
Joseph Henry Laboratories\\
Princeton University\\
Princeton, New Jersey 08544
\end{center}
\vspace{7mm}
\begin{center}
{\large Abstract}
\end{center}
\noindent 
In this article  old and new relations between gauge fields and strings are 
discussed. We add new arguments that the Yang -Mills theories must be described
by the non-critical strings in the five dimensional curved space. 
The physical meaning of the fifth  dimension is that of the renormalization scale 
represented by the Liouville field.
We analyze the meaning of the zigzag symmetry  and show that it is likely to 
be present if there is a minimal supersymmetry on the world sheet. 
We  also present  the new string backgrounds which may be relevant for the 
description of the ordinary bosonic  Yang-Mills theories. 
The article is written on the occasion of the 40-th anniversary of the IHES.
\vspace{7mm}
\begin{flushleft}
September 1998

\end{flushleft}
\end{titlepage}

\section{Introduction}

Very often a source of strong poetry and strong science is a good metaphor.
My favorite one is Plato's cave: the parable of the men sitting in a dark
cave, watching the moving shadows on its wall. They think that the shadows
are ''real'' and not just projections of the outside world.

It seems to me that the latest stages of the ongoing struggle to understand
interactions of elementary particles create a picture stunningly close to
this parable.

In this article I will try to summarize these latest developments, adding
some new conjectures and results. Of course it should be remembered that the
work is far from finished and many surprises lie ahead.

It was understood long ago that compactness of the gauge group is likely to
lead to the collimation of the flux lines and thus to quark confinement.
This was first discovered by K. Wilson [1 ] in the strong coupling limit of
lattice gauge theories. The next step was the proof of quark confinement for
the compact abelian gauge groups based on instanton mechanism [2,3 ] . These
theories arise from the non-abelian ones after partial symmetry breaking.
Random fields of instantons (which are magnetic monopoles in 3d and closed
rings of monopole trajectories in 4d) prevent propagation of charged objects
and collimate the flux lines. This picture received a physical
interpretation in terms of the ''dual superconductors'' [4,5].

This approach is not directly extendable to the nonabelian theories and only
partial progress was made in this field until recently. The new development
is based on the old idea that gauge theories must allow an exact description
in terms of strings representing the Faraday flux lines and also on a new
concept of D-branes.

Here is a brief recent history of these developments. Some years ago I
became convinced that the string theory, describing gauge fields is a
Liouville theory with the curved fifth dimension. The guiding principle
corresponding to the gauge symmetry was claimed to be the zigzag invariance
of the boundary action. That insured that only vector states survive at the
open string boundary. Thus the problem was reduced to the solution of the
nonlinear 2d sigma model in which the string tension may become either zero
or infinite (these cases are related by T-duality and equally well satisfy
the zigzag symmetry; the choice between them requires more dynamical
information and will be discussed below). This problem still remains
central. These results were summarized in [6 ].

Meanwhile there has been an apparently unrelated development. In a seminal
paper [30] Polchinski introduced the idea that D-branes in the type II
critical strings can be regarded as gravitational solitons previously
discovered by Horowits and Strominger [24]. After that in a remarkable paper
[7 ] Ig. Klebanov compared the absorption of external particles by the
3-branes, viewed as maximally supersymmetric Yang -Mills theory, with the
picture of 3-branes, viewed as a gravitational soliton. He noticed that the
correspondence should exist in the large N limit and for the large t'Hooft
coupling (since in this case the sigma model corrections are small), and
confirmed this by the explicit computations. In [31] some of the Yang-Mills
correlators were computed by this methods. Later in an insightful work J.
Maldacena [8] noticed that the metric relevant to these calculations is that
of $AdS_{5}\times S^{5},$ conjectured that the super Y-M theory corresponds
to the full string theory in this background and pointed out that the
distance to the D-brane can be identified with the Liouville field of the
previous approach [6]. This synthetic point of view was used in [9, 10 ] to
formulate the rules for calculating correlators and anomalous dimensions in
the above theory. After that this side of the subject received enormous
development.

In this paper we concentrate on the more difficult but physically important
aspects, not related to the supersymmetry. We shall begin our discussion by
reviewing the basic properties of the non-critical strings and random
surfaces.

\section{The non-critical string and the Liouville dimension..}

The propagating strings sweep out world surfaces. The quantum string theory
is essentially the theory of these random surfaces. The classical action for
the purely bosonic string is given by [11, 12] 
\begin{equation}
S=\int \sqrt{g(}g^{ab}\partial _{a}x^{\mu }\partial _{b}x^{\mu }+\lambda
)d^{2}\xi
\end{equation}

with $g_{ab}(\xi )$ being an independent metric. Here we parametrize the
random surface by $x^{\mu }=x^{\mu }(\xi ^{1},\xi ^{2}),\mu =1,...D,$ where $%
D$ is the dimension of the ambient space. This is the most general action
which is parametrically invariant and contains the minimal number of
derivatives. The metric seems to disappear from (1) if we chose the
conformal gauge, $g_{ab}=e^{\varphi }\delta _{ab}$ . However this is not so
because of the quantum anomaly, and the effective action describing the
string in the conformal gauge was shown to be [13] 
\begin{equation}
S=\int d^{2}\xi [\frac{26-D}{96\pi }(\partial \varphi )^{2}+\lambda
e^{\varphi }+\frac{1}{2}(\partial x^{\mu })^{2}]
\end{equation}

Here the constant $\lambda $ must be adjusted in order to reach the
continuum limit.

Still this is not the end of the story. One has to find a way of quantizing
the $\varphi $- field consistently with the conformal invariance. The reason
why the conformal symmetry on the world sheet is sacred is that it is a
remnant of the general covariance in the conformal gauge, and thus any
quantization scheme preserving the above covariance must be conformally
invariant. On the other hand, any standard cut-off in the $\xi $ - space
would destroy this property. The way to overcome this difficulty suggested
in [14, 15] was essentially to retain the standard cut-off but to modify the
action in such a way that the conformal symmetry is intact. In the
simplified case of models with the central charge $c\leq 1$ one has to
consider the action of the form 
\begin{equation}
S=\int d^{2}\xi \{(\partial \varphi )^{2}+(\partial x)^{2}+T(\varphi )+\Phi
(\varphi )R_{2}\}
\end{equation}

where $R_{2}$ is the fixed curvature of the world sheet and the functions $T$
and $\Phi $ are determined by conformal invariance. It has been shown in
[16] along the general lines of [17] that the above invariance requires
minimization of the effective action 
\begin{equation}
W[T]=\int_{0}^{\infty }e^{-2\Phi }[(\frac{dT}{d\varphi })^{2}-2T^{2}+V(T)]
\end{equation}

where $\Phi =b\varphi $ and $1+12b^{2}=26$ $-c$ ; $V(T)$ is some potential.
More generally one may need other fields to be included into the action
(this does not happen for the minimal models).

The partition function $Z(\lambda )$ of the string is given by the value of
this action on the classical solution 
\begin{equation}
Z(\lambda )=W[T_{cl}(\varphi )];T_{cl}(0)=\lambda
\end{equation}

In general the partition function satisfies the Hamilton - Jacobi equations
which can be interpreted as a non-linear renormalization group [18]. The
partition function has a singularity in $\lambda $ which is related to the
tachyonic nature of the field $T$ . Indeed if we replace $T$ by some massive
field $\Psi $ in (4) and neglect the potential the classical solution will
have the following asymptotic behaviour 
\begin{equation}
\Psi (\varphi )\sim e^{b\varphi -\sqrt{b^{2}+M^{2}}\varphi };\varphi
\rightarrow \infty .
\end{equation}

We conclude that there are three types of the large $\varphi $ behaviour.
First of all there are stable massive modes. The action is analytic in their
initial data. Next there could be ''good'' tachyons with $-b^{2}$ $\leq
M^{2}\leq 0.$ These tachyons begin to grow and then condense after being
stopped by the potential $V(T)$ . This is what happens in the models with $%
c\leq 1$ and generates the singularity of $Z(\lambda )$. These models are
perfectly healthy in spite of the tachyon. Finally, if $M^{2}\leq -b^{2}$ we
have a case of ''bad'' tachyon. The theory becomes unstable and the
effective action looses its positivity because of the oscillations of the
corresponding modes. This happens when $c>1.$

It was suggested in [6] that this problem can be overcome if we assume that
the $(\varphi ,x)$ geometry for $c>1$ is not flat anymore. Since the $(x)$
geometry must remain flat, the most general form of action takes the form 
\begin{equation}
S=\int (\partial \varphi )^{2}+a^{2}(\varphi )(\partial
x)^{2}+\sum_{n}\lambda _{n}(\varphi (\xi ))V_{n}(\xi )
\end{equation}

Here the vertex operators $V_{n}$ are describing the modes with $M^{2}\leq 0$%
, since only these modes have the tendency to condense. The function $%
a(\varphi )$ representing the running string tension and the couplings $%
\lambda _{n}(\varphi )$ must be determined from the condition of conformal
invariance on the world sheet. The effective action for the tachyon is now
modified due to the non-flatness of the $(\varphi ,x)$ geometry. It has the
form 
\begin{equation}
W=\int d\varphi e^{-2\Phi }a^{D}(\varphi )[(\frac{dT}{d\varphi }%
)^{2}-2T^{2}+...]
\end{equation}

The dots include the action for $a$ and $\Phi $ which we will discuss later
and also possible higher derivative terms.The evolution of the tachyon now
takes place in the ''expanding universe'' (with $\varphi $ playing the role
of the euclidean time). It is well known that expansion tends to moderate
instabilities. For example Jeans instability in the flat space is
exponential, while in the Friedmann universe it is power-like. Something
similar may happen in our case. If we assume that there is a solution ,
minimizing the above action, of the ''big bang'' type 
\begin{eqnarray}
a(\varphi ) &\sim &\varphi ^{\alpha } \\
\Phi (\varphi ) &\sim &\log \varphi
\end{eqnarray}

The tachyon will behave as $T(\varphi )\sim \varphi ^{\gamma }$ with some
real $\gamma $. This type of solutions exist in the one loop approximation.
Whether they can be promoted to exact conformal field theories is a
difficult open question (which is somewhat more tractable in supersymmetric
cases as we will discuss later). We use it here as an illustration of a
possible scenario. Namely, the tachyon grows and tends to a constant, as it
does at $c=1.$ As a result we obtain a theory of random surfaces in which
the string tension $\sigma =a^{2}$ and the cosmological constant $\lambda =T$
are related by the scaling law $\sigma \sim \lambda ^{\frac{2\alpha }{\gamma 
}}$. Of course this illustration must be taken with a grain of salt since it
is based on the low energy effective action. It well may happen that in
order to stabilize a random surface one needs supersymmetry on the world
sheet. At least, as we will argue in the next sections, this supersymmetry
is needed to describe gauge theories. Nevertheless the bosonic theory is an
important preliminary step.

Let us summarize what we have learned. First, due to the quantum anomalies,
the target space of the $D$ -dimensional random surface has $D+1$
dimensions. This happens because renormalization changes the classical
action of the random surface in the following way 
\begin{equation}
\sqrt{g}g^{ab}\partial _{a}x\partial _{b}x\Rightarrow (\partial \varphi
)^{2}+a^{2}(\varphi )(\partial x)^{2}+...
\end{equation}

All the background fields must be determined from the condition of conformal
invariance. The extra dimension, grown by the quantum effects, comes out of
the conformal factor of the world sheet metric. The background fields of the
original, $D$ - dimensional random surface, play the role of the initial (or
,better to say, boundary) data for the $D+1$ dimensional effective action.
This action, considered as a function of the initial data is equal to the
partition function of the non-critical string.

\section{Gauge fields and zigzag symmetry.}

The loop equations provide a bridge between gauge theories and strings
although this bridge is not completely safe. They are written for the
functional $W(C)$ which is the expectation value of the parallel transport
around the loop $C$ , calculated with the Yang-Mills action. Their role is
to translate the Schwinger-Dyson equations for the Yang -Mills fields into
the language of loops. The role of string theory is to solve these equations
in terms of random surfaces. The equations have the form [19, 20 ] 
\begin{equation}
\frac{\partial ^{2}W(C)}{\partial x^{2}(s)}=\int \delta (x(s)-x(u))\frac{dx}{%
ds}\frac{dx}{du}W(C_{1})W(C_{2})
\end{equation}

Here the operator on the left hand side is defined as a coefficient in front
of the $\delta $- function (the contact term) in the second variational
derivative, the $\delta $- function on the right hand side is nonzero if the
disc bounded by $C$ is pinched, and the contour splits into two parts, $%
C_{1} $ and $C_{2}.$ The gauge field -string equivalence means that there is
a string action, containing the Liouville field and maybe other fields,
which solves this equation in the following sense 
\begin{equation}
W(C)=\int Dx(\xi )D\varphi (\xi )e^{-S(\varphi ,x)}
\end{equation}

Here the $\xi $ space is assumed to be a disk or half-plane. At the boundary
of this disc , we impose the Dirichlet conditions 
\begin{equation}
x\mid _{\partial D}=x(s);\varphi \mid _{\partial D}=\varphi _{*}
\end{equation}

The first part of this choice is obvious - we just map the world sheet disk
into a disk in the target space, bounded by our loop. The boundary
conditions for the Liouville $\ $field are much more subtle. It is important
to realize that they can not be chosen arbitrarily. The conformal invariance
of the world sheet imposes very strong constraints and in particular fixes
the value of $\varphi _{*}$ if the allowed value exists at all. The second
strong requirement is the zigzag symmetry. Let us discuss this requirement.
The Wilson loop is invariant under reparametrizations of the contour 
\begin{equation}
x(s)\Rightarrow x(\alpha (s))
\end{equation}

Since the string theory is invariant under diffeomorphisms of the world
sheet, the above symmetry is automatic if $\alpha (s)$ is a diffeomorphism,
which means that $\frac{d\alpha }{ds}>0.$ However the basic feature of the
Wilson loop is the invariance under zigzags, when the last condition is
dropped. We shall see now that in string theory the necessary condition for
such an extended reparametrization invariance is $a(\varphi _{*})=0$ or $%
\infty .$

Let us discuss the first requirement. The reason why the generic Dirichlet
condition is impossible in the curved space is that the D-brane defined by
it will be subjected to gravitational forces tending to change its shape.
This can be seen as following. Suppose that we have a 2d sigma model with
the target space metric $G_{MN}(y)$ where $y=(x,\varphi ).$ Let us try to
impose the condition at the boundary of the world sheet disc $\varphi \mid
_{\partial D}=\varphi _{*}(x)$. The effective action acquires a boundary
term 
\begin{equation}
W_{B}=\int d^{4}xe^{-\Phi (x,\varphi _{*}(x))}\sqrt{G(x,\varphi _{*}(x))}
\end{equation}

The function $\varphi _{*}(x)$ must be chosen so as to minimize this action
(or the modified action when the higher derivative terms are added). This
condition defines a possible shape of the D-brane. In our case it gives 
\begin{equation}
\frac{\partial }{\partial \varphi }(e^{-\Phi }a^{4})=0
\end{equation}

If, as it often happens, the particular solution does not satisfy this
condition, the D-brane will be carried either to zero or to infinity where a
separate consideration is needed. Namely, the zigzag invariance of the
Wilson loop is reflected in the properties of the algebra of boundary
operators. As usual, the open string amplitudes can be expressed as 
\begin{equation}
A(p_{1}...p_{n})=\int ds_{1}...ds_{n}\int
Dx(s)W[x(s)]V_{p_{1}}(x(s_{1}))...V_{p_{n}}(x(s_{n}))
\end{equation}

The vertex operators $V_{p}(x)$ are in general exponentials $e^{ipx}$
multiplied by the product of different derivatives of $x(s).$ The usual
reparametrization invariance is translated into the Virasoro conditions,
satisfied by those vertex operators. That still leaves an infinite number of
possible vertex operators, representing the states of the open string. When $%
W[x(s)]$ is zigzag invariant the situation is different. In this case the
only allowed vertex operator is the vector one 
\begin{equation}
V_{\alpha }(p)=\frac{dx_{\alpha }}{ds}e^{ipx(s)}
\end{equation}
since otherwise the integrand in (18) will loose its zigzag symmetry. That
was our main conclusion in [6]: {\em the} {\em zigzag symmetry is equivalent
to the finiteness of the algebra of boundary operators which contains only
vector vertices.}

The other way to put it is to say that with the above boundary conditions
the open string contains only gluons in its spectrum. If there are other
fields in addition to the Yang -Mills, their vertex operators must be added
at the boundary. In general {\em the spectrum of the boundary states is the
spectrum of the corresponding field theory.}

It would be nice to be able to check directly that the loop equations are
satisfied by our ansatz.. Some relevant methods were discussed in [6, 21,
22] but they are still insufficient. However this check may be not so vital.
The loop equations are bound to be satisfied by the following argument. Our
open string theory has only vector particles and only positive norms.
Therefore it must be the Yang -Mills theory. This reminds of the standard
deduction of the Einstein theory from the existence of massless tensor
particles.

If try to impose the Dirichlet boundary conditions for the $\varphi $ -field
at the finite value of the string tension $a^{2}(\varphi _{*})$the whole
tower of the open string states will be present and the zigzag symmetry will
not be there. That leaves us with two options, either to place the gauge
theory at the horizon, $a(\varphi _{*})=0,$ or at infinity, $a(\varphi
_{*})=\infty .$ It seems that both choices will provide us with the zigzag
symmetry and thus are equivalent. The first one, preferred in [6],
eliminates the $(\partial x$ $)^{2}$ term from the action. However the
presence of the Wilson loop generates the term $B_{\mu \nu }(x)\epsilon
^{ab}\partial _{a}x^{\mu }\partial _{b}x^{\nu }$ in the string action which
becomes dominant in this case. This term has the interpretation as a color
electric flux [6]. However, it is difficult to find its explicit form and
for that reason in the papers [9, 10, 23 ] a simpler choice was suggested ,
placing the gauge theory at infinity, also leading to the zigzag symmetry,
has been . In this case the $B_{\mu \nu }-$ term is negligible since $%
a(\varphi _{*})=\infty $ and we can concentrate on a simpler set of
background fields. Let us give a simple although non-rigorous argument which
demonstrates the presence of the zigzag symmetry in the case of the infinite
tension. Consider some massive mode of the open string, $\Psi (\varphi
_{*},x).$ The quadratic part of the boundary effective action has the form 
\begin{equation}
W_{B}\sim \int dx[a^{-2}(\varphi _{*})(\frac{d\Psi }{dx})^{2}+M^{2}\Psi ^{2}]
\end{equation}

If $a(\varphi _{*})=\infty $ the first term drops out and we conclude that
minimization of the effective action requires $\Psi =0.$

On the other hand the massless mode can have an arbitrary $x$ dependence.
This is just what is desirable for the description of the gauge theory,
since we do not want to restrict the momentum carried by the vertex
operators. It must be said however that the choice of boundary conditions
requires further clarification which must be based on the dynamical
arguments. This has not been done.

Another important conclusion is that the tachyons will condense and,
generally speaking, will be present at the boundary thus violating the
zigzag symmetry. We shall get rid of them in the next section. Of course
these arguments must be improved by considering conformal field theory
represented by the above non-linear sigma model. This also has not been done
so far.

\section{Elimination of the boundary tachyon.}

We saw that in the purely bosonic Liouville theory we expect that some
scalar tachyonic modes will penetrate the boundary. Also, the Wilson loops
and related quantities have some awkward divergences which made me suspect
in [6] that supersymmetry on the world sheet may be helpful in comparing
gauge fields and strings. In this section we will present a simple argument
showing that in the Liouville theory with minimal supersymmetry on the world
sheet and non-chiral GSO projection the boundary tachyon is absent and the
zigzag symmetry is to be expected. This theory is purely bosonic in the
target space and, according to the above arguments, must describe the pure
Yang-Mills theory. In the theories of this kind the world sheet is
parametrized by the coordinates $(\xi ^{1},\xi ^{2},\theta ^{1},\theta ^{2})$%
, where $\theta $ are anticommuting. The boundary of the world sheet is
parametrized by one bosonic and one fermionic variable, $(s,\vartheta ).$ It
is well known that the formulas of the bosonic non-critical string theory
are almost unchanged in the fermionic case if written in this superspace. In
particular the loops and the loop equations are changed as follows 
\begin{eqnarray}
\int A_{\mu }\frac{dx^{\mu }}{ds}ds &\Rightarrow &\int dsd\vartheta A_{\mu
}(x(s,\vartheta ))D_{\vartheta }x^{\mu } \\
D_{\vartheta } &=&\frac{\partial }{\partial \vartheta }+\vartheta \frac{%
\partial }{\partial s}
\end{eqnarray}

and the only change in the loop equation is the replacement of the $s$-
derivatives by the $D_{\vartheta }.$ The same minimal change takes place in
the non-critical string action (11) . It retains the same form in terms of
superfields, with the $\xi -$ derivatives being replaced by the $\theta -$
derivatives. The above generalized Wilson loop describes the spin one half
test particle.

There are, however, some important differences from the bosonic case. First
of all the critical dimension in this string is $D_{cr}=10$ and not 26. A
more subtle difference is that to be consistent the theory requires the GSO
projection. This projection is a summation over the spin structures, that is
as we go around a cycle or an injection point, the world sheet fermions may
or may not change sign. We must sum over the possibilities since otherwise
the modular invariance and unitarity will be broken. In the critical
superstrings the standard prescription is to sum over the signs of the left
and right fermions independently. That adds to the operator algebra left and
right spin operators which transform as space-time fermions and generate 10d
supersymmetry. It also truncates the algebra of bosonic vertices by dropping
those which contain odd number of either left or right world sheet fermions.

In the non-critical strings with the variable string tension there is a
slightly different way to make the GSO projection. Namely it is consistent
to assume that the projection is non-chiral or that, as we go around the
cycle, left and right fermions change in the same way. This will again
introduce the spin operators but this time they will transform as a product
of the space time fermions and will correspond to the Ramond-Ramond bosons.
There also will be a truncation of the algebra of vertex operators, but this
time only the total number of the world sheet fermions will have to be even.
As a result in the closed string sector we have a tachyon described by a
vertex operator 
\begin{equation}
V(p)=\int d^{2}\xi d^{2}\theta \chi _{p}(\varphi (\xi ,\theta ))e^{ipx(\xi
,\theta )}
\end{equation}

where the function $\chi _{p}(\varphi )$ is determined from the condition of
conformal invariance on the world sheet (which reduces to the Laplace
equation in the metric (7 ) in the weak coupling limit. Since this
expression is even in $\theta $ it will be also even in the world sheet
fermions after the $\theta $- integration is performed. Thus this vertex is
allowed by the non-chiral GSO projection. Presumably this tachyon should be
of the ''good'' variety and peacefully condense in the bulk. The important
thing is not to have a boundary tachyon which would spoil the loop equation.
The GSO projection takes care of that. The boundary tachyon vertex has the
form 
\begin{equation}
V_{B}(p)=\int dsd\vartheta \chi _{p}^{(B)}(\varphi (s,\vartheta ))e^{ipx}
\end{equation}

and clearly contains odd number of fermions. It is eliminated by the GSO
projection and we are left with the purely vector states surviving at the
boundary.

We conclude that the supersymmetric Liouville theory with the non-chiral GSO
projection is likely to describe the bosonic Yang-Mills theory. An important
feature of this theory is the presence of the bosonic Ramond-Ramond fields.
These fields can create new conformal points. This happens in statistical
mechanics where the imaginary magnetic field in the 2d Ising model (which is
the simplest example of the RR-field) drives the theory to a new so-called
Lee- Yang critical point. We shall see that something similar happens in our
case. In the next section we discuss the extent to which we can believe in
the existence of the conformal theories with the curved geometry in the 5d
Liouville space.

\section{ Cosmology inside hadrons (conformal symmetry on the world sheet).}

As we already discussed, the conformal symmetry of our sigma model is
absolutely necessary for the consistency of the whole approach. According to
[17], the background fields providing conformal invariance must satisfy the
so-called $\beta -$ function equations. We know their explicit form only in
the low energy approximation, which generally speaking is not sufficient for
our purposes. This is the main obstacle for our approach. Still in some
cases (which we discuss below) this approximation is good enough and also it
often gives a correct qualitative picture. When the RR-fields are present
the $\beta -\ $function equations take the form 
\begin{equation}
R_{AB}=\nabla _{A}\nabla _{B}\Phi +e^{-\Phi }\Theta _{AB}
\end{equation}

where $\Theta _{AB}$ is the energy-momentum tensor for the RR-fields,
indices $A$ and $B$ include the Liouville direction and we changed the
normalization of the dilaton 
\begin{equation}
-2\Phi \Rightarrow \Phi
\end{equation}

There is also the condition balancing the central charge 
\begin{equation}
R-2\nabla ^{2}\Phi -(\nabla \Phi )^{2}+\frac{10-D}{2}=0
\end{equation}

where $D$ is the total number of dimensions. The RR -fields appear as
antisymmetric tensors of different ranks.

If we denote the rank $p+1$ field by $C_{p+1}$ the energy-momentum tensor is
expressed in terms of the field strength $F_{p+2}=dC_{p+1}.$ The RR fields
satisfy the equation 
\begin{equation}
d^{+}F_{p+2}=0
\end{equation}

Their energy momentum tensor is given by 
\begin{equation}
\Theta _{AB}=F_{A...}F_{B...}-\frac{1}{2(p+2)}G_{AB}F_{...}F_{...}
\end{equation}

where the dots mean the indices contracted with the metric $G$. Let us look
for the solution which leaves the $x-$ space flat. That means that the
metric takes the form (11). To incorporate some interesting examples in
which extra fields are added to the Yang-Mills we shall consider the
following ansatz 
\begin{eqnarray}
ds^{2} &=&(d\varphi )^{2}+a^{2}(\varphi )(dx)_{p+1}^{2}+b^{2}(\varphi
)(dn)_{q+1}^{2} \\
C_{p+1} &=&c(\varphi )dx^{1}...dx^{p+1}
\end{eqnarray}

where $n$ -represents the additional q+1 dimensional sphere and only the $%
p+1 $ form was assumed to be non-zero. This is the only possibility
consistent with the symmetries of the $x-$ space apart from switching also
the zero form, corresponding to the theta terms. This class of solutions has
been considered in the past as a description of the p-branes [24 ] (on which
we will comment in the next section) and in the case of the Minkowskian
signature as a cosmological models (see e.g. [ 25]). However we have to
concentrate on some special cases . The equation (28 ) gives 
\begin{equation}
\frac{dc}{d\varphi }=N\frac{a^{p+1}}{b^{q+1}}
\end{equation}
where $N$ is the integration constant. The equations (25) and (27) in this
ansatz take the form 
\begin{equation}
\stackrel{\cdot \cdot }{\Phi }+(p+1)a^{-1}\stackrel{\cdot \cdot }{a}%
+(q+1)b^{-1}\stackrel{\cdot \cdot }{b}=N^{2}e^{-\Phi }\frac{1}{b^{2q+2}}
\end{equation}
\begin{equation}
a^{-1}\stackrel{\cdot \cdot }{a}+pa^{-2}\stackrel{\cdot }{a}^{2}+a^{-1}%
\stackrel{\cdot }{a}\stackrel{\cdot }{\Phi }+(q+1)a^{-1}b^{-1}\stackrel{%
\cdot }{a}\stackrel{\cdot }{b}=N^{2}e^{-\Phi }\frac{1}{b^{2q+2}}
\end{equation}
\begin{equation}
b^{-1}\stackrel{\cdot \cdot }{b}+qb^{-2}\stackrel{\cdot }{b}^{2}+b^{-1}%
\stackrel{\cdot }{b}\stackrel{\cdot }{\Phi }+(p+1)a^{-1}b^{-1}\stackrel{%
\cdot }{a}\stackrel{\cdot }{b}-qb^{-2}=-N^{2}e^{-\Phi }\frac{1}{b^{2q+2}}
\end{equation}
\begin{equation}
\stackrel{\cdot \cdot }{\Phi }+(p+1)a^{-1}\stackrel{}{\stackrel{\cdot }{a}%
\stackrel{\cdot }{\Phi }}+(q+1)b^{-1}\stackrel{\cdot }{b}\stackrel{\cdot }{%
\Phi }+\stackrel{\cdot }{\Phi }^{2}=\frac{10-d}{2}-N^{2}(p-q+1)e^{-\Phi }%
\frac{1}{b^{2q+2}}
\end{equation}
Here $d$ is the total dimensionality, $d=p+q+3.$ In general these equations
have power- like , the ''big bang'' solutions. At best they can be taken as
a hint that there exist the corresponding nonlinear sigma model with
conformal invariance. There are however some cases in which the equations
can be taken more seriously. These are the cases in which the curvature of
the 5d Liouville space is constant. According to [8] the constant curvature
solution correspond to the conformal symmetry in the target space and thus
are more tractable. A look at the above equations shows that there are at
least two possibilities to have such solutions. First let us take $%
d=p+q+3=10 $ and $p-q+1=0.$ According to (36) the driving force for the
dilaton vanishes and it can be set to a constant. We then easily solve the
remaining equations by setting $b(\varphi )=const$ and $a(\varphi
)=e^{\alpha \varphi } $ $.$ In this case $p=3$ , $q=4$ and thus we are
dealing with the space $L_{5}\times S^{5}$ where $L_{5}$ is the 5d
Lobachevsky space (the Minkowskian version of which is called $AdS)$and $%
S^{5}$ is a 5d sphere. All parameters are fixed by the above equation and
have the following orders of magnitude 
\begin{equation}
e^{-\Phi }\sim g_{s}^{2};\frac{1}{b^{8}}\sim (g_{s}N)^{2};\alpha ^{2}\sim 
\frac{(g_{s}N)^{2}}{b^{10}}\sim (g_{s}N)^{-\frac{1}{4}};
\end{equation}
where $g_{s}$ is the closed string coupling constant.

This is of course the standard 3-brane solution [24] rescaled near the brane
and written in the Liouville gauge. It describes the maximally
supersymmetric Yang-Mills theory [7, 8]. We see that if we take $%
N\rightarrow \infty ,$ while keeping $g_{s}N$ fixed we can neglect the
closed string loops. It is also possible to neglect the sigma model
corrections if the curvature of this solution, which is $\sim \alpha ^{2}$
is small. Hence the whole construction is trustworthy provided that $%
g_{s}N>>1.$ This is the much explored case first analyzed in [7] . We shall
not discuss it further except noticing that the real remaining challenge in
this case just as in the other theories we are discussing here is to promote
the above solution to the complete conformal field theory.

Let us turn to another constant curvature case which is less reliable then
the above but much more physically interesting (the usual dichotomy). Namely
let us consider the pure non-supersymmetric Yang-Mills theory . That means
that this time we have no $n-$ field and have to set $q=-1$ . We can keep
the dilaton from running if we chose the string coupling so as to set the
right hand side of the eq. (36) to zero. In this case 
\begin{equation}
g_{s}N=\sqrt{\frac{8-p}{2(p+1)}}
\end{equation}
After that the equations are easily solved by $a(\varphi )=e^{\alpha \varphi
}$ and we have a constant curvature solution once again. It describes a
conformal fixed point of the Yang -Mills theory. Unfortunately we can't be
certain that the sigma model corrections (which are of the order of unity)
will not destroy this solution. Usually however they don't , just
renormalizing parameters, but not spoiling conformal symmetry. If this is
the case, we come to the conclusion that strong interactions at small
distances are described not by asymptotic freedom but by a conformal field
theory, a possibility dreamt of in [26,27 ]. Once again, the exact solution
of the sigma model is needed to prove or disprove this drastic statement.

If there is no fixed point we must recover the asymptotic freedom from our
string theory. That means that the dilaton and the curvature will be the
functions of $\varphi .$ The ''conformal time''$\tau =$ $\int \frac{d\varphi 
}{a(\varphi )}$ defines an effective scale (perhaps related to the comments
in [28 ]). Therefore the asymptotic freedom will manifest itself in the
following relation 
\begin{equation}
g_{s}N\sim g_{YM}^{2}N\sim e^{-\frac{\Phi }{2}}\sim \frac{1}{\log (\tau )}
\end{equation}
In checking this relation the one loop approximation in the sigma model is
even less adequate then in the previous case, since the YM coupling is small
and the curvature of the Liouville space is large. However the problem is
not hopeless. We need the solution of the sigma model with the large but
slowly varying curvature. That means that we have to account the curvature
terms in all orders but can neglect their derivatives (analogously to what
is done in the case of the Born-Infeld action for the open strings [29 ] ).
There are reasons to believe that the above sigma models with constant
curvature are completely integrable. Thus we may hope to find the complete
solution of the gauge fields -strings problem and perhaps even to discover
experimental manifestations of the fifth (Liouville) dimension.

\section{Acknowledgments}

I am delighted to have an opportunity to express my gratitude to the IHES,
the best place on earth to do science, surrounded by the top minds and
beautiful gardens.

I also would like to thank T. Damour, V. Kazakov, I. Klebanov, I. Kostov and
J. Maldacena for very important discussions and suggestions.

This research was supported in part by the National Science Foundation under
Grant No. PHY96-00258.

\newpage REFERENCES

[1] K. Wilson Phys. Rev.D10(1974)2445

[2] A. Polyakov Phys.Lett.59B(1975)82

[3] A.Polyakov Nucl. Phys. B120(1977)429

[4] S. Mandelstam Phys. Rep. 23C(1976)245

[5] G. t'Hooft in High Energy Phys., Zichichi editor, Bolognia (1976)

[6] A. Polyakov HEP-TH/9711002

[7] I. Klebanov Nucl. Phys. B496(1997)231

[8] J. Maldacena HEP-TH/9711200

[9] S. Gubser I. Klebanov A. Polyakov HEP-TH/9802109

[10] E. Witten HEP-TH/9802150

[11] L. Brink P. di Vecchia P. Howe Phys. Lett. 63B(1976)471

[12] S. Deser B. Zumino Phys. lett. 65B(1976)369

[13] A. Polyakov Phys. Lett. 103B(1981)207

[14] T. Curtright C. Thorn Phys. Rev. Lett. 48(1982)1309

[15] J. Gervais A. Neveu Nucl. Phys. B199(1982)59

[16] J. Polchinski Nucl. Phys. B346(1990)253

[17] C. Callan E. Martinec M. Perry D. Friedan Nucl. Phys. B262(1985)593

[18] A. Polyakov Proceedings of Les Houches (1992)

[19] A. Polyakov Nucl. Phys. B164(1980)171

[20] Y. Makeenko A. Migdal Nucl. Phys. B188(1981)269

[21] H. Verlinde HEP-TH/9705029

[22] A. Migdal Nucl. Phys. Proc. Suppl. 41(1995)51

[23] J. Maldacena HEP-TH/9803002

[24] G. Horowitz A. Strominger Nucl. phys. B360(1991)197

[25] A. Lukas B. Ovrut D. Waldram HEP-TH/9802041

[26] K. Wilson Phys. Rev.179(1969)1499

[27] A. Polyakov ZHETF 59 (1970) 542 ; Pisma v ZHETF 12(1970)538

[28] A. Peet J. Polchinski HEP-TH/9809022

[29] E. Fradkin A. Tseytlin Phys. Lett.B178(1986)34

[30] S. Gubser I. Klebanov HEP-TH/9708005

\end{document}